\documentclass[aps, prl, superscriptaddress, twocolumn, showpacs]{revtex4-1}
\usepackage[utf8]{inputenc}
\usepackage{bm}
\usepackage{mathtools}
\usepackage{bbold}
\usepackage{amsmath}
\usepackage{amssymb}
\usepackage{color}
\usepackage[dvipsnames]{xcolor}
\usepackage{multirow}
\usepackage{comment}
\usepackage{cancel}
\usepackage{float}
\usepackage{color}
\usepackage[dvipsnames]{xcolor}
\usepackage{epstopdf}
\usepackage{amsmath}
\usepackage{braket}
\usepackage{amsthm}
\usepackage{amssymb}
\usepackage{amsfonts}
\usepackage{graphicx,subfigure}
\usepackage{txfonts}
\usepackage{ulem}
\usepackage{mathtools}
\usepackage{bm}
\usepackage{bbm}
\usepackage{hyperref}
\usepackage{natbib}
\newcommand{\Eins}{\ensuremath{\mathbbm 1}}
\newcommand{\vect}[1]{\boldsymbol{#1}}

\begin{document}

\newcommand{\INO}{Istituto Nazionale di Ottica, Consiglio Nazionale delle Ricerche (INO-CNR), Largo Enrico Fermi 6, 50125 Firenze, Italy}
\newcommand{\LENS}{European Laboratory for Nonlinear Spectroscopy (LENS), Via N. Carrara 1, 50019 Sesto Fiorentino, Italy}
\newcommand{\UNIFI}{Physics Department, University of Florence, Via G. Sansone 1, 50019 Sesto Fiorentino FI, Italia
}
\newcommand{\be}{\begin{equation}}
\newcommand{\ee}{\end{equation}}

\title{Rotation Sensing via Josephson-frequency Splitting in a Toroidal Superfluid}

\author{Giulio Nesti}
\affiliation{\INO}
\affiliation{\LENS}
\author{Luca Pezz\`e}
\affiliation{\INO}
\affiliation{\LENS}
\date{\today}
	
\begin{abstract}
We show that a toroidal superfluid interrupted by $n$ tunneling barriers realizes a compact Josephson gyroscope with an $n$-enhanced response to rotation. 
In the small-amplitude regime, we derive analytically the normal mode spectrum of the coupled population–phase oscillations. 
In the absence of rotation, pairs of modes are degenerate: a finite angular velocity $\Omega$ lifts this degeneracy through a Doppler shift, producing a frequency splitting that grows linearly with both $\Omega$ and $n$. 
Full numerical simulations confirm this prediction and reveal long-lived two-frequency beatings, in sharp contrast with the monochromatic Josephson oscillations of the nonrotating system. 
These beatings provide a direct rotation signal with estimation uncertainty scaling as $\Delta\Omega\sim n^{-3/2}$, while remaining robust against imperfections and dynamical excitations.
These results identify multi-junction toroidal superfluids as scalable, micrometer-size rotation sensors compatible with current experimental platforms.
\end{abstract}

\maketitle

{\it Introduction.---}
Ultracold atoms in toroidal traps~\cite{PoloPR2025} provide a highly controllable platform for studying persistent currents~\cite{RyuPRL2007,KumarNJP2016,KumarPRA2018,CaiPRL2022,DelPacePRX2022}, weak links~\cite{WrightPRL2013,RyuPhys2013,RyuNature2020}, and superfluid transport~\cite{RamanathanPhys2011,JendrzejewskiPRL2014,EckelPRX2014}.
Recent advances in optical trapping have enabled realization of smooth ring confinements~\cite{DelPacePRX2022}, multiple stable tunnel barriers~\cite{RyuNature2020,PezzeNATCOMM2024}, and controllable defects~\cite{XhaniImpurities2025}, opening new opportunities for both fundamental studies of superfluidity~\cite{GanPRR2026,TüzemenPRR2026,BorysenkoPRA2025,Hernández-RajkovNATPHYS2024,ChenPRR2025,CiszakARXIV2025,NestiARXIV2025} and compact rotation sensing~\cite{MartiPRA2015,WoffindenSCIPOST2023}.
Ring-shaped ultracold gases address a regime complementary to superfluid-helium gyroscopes~\cite{PackardPRB1992,SchwabNATURE1997,SatoRPP2012} and Sagnac matter-wave interferometers~\cite{GustavsonPRL1997,LenefPRL1997,BarrettCRP2014,DuttaPRL2016,GautierSCIADV2024}.
While the latter achieve high precision by maximizing the enclosed area, microscopic superfluid rings trade area for compactness, scalability, and potential integration into portable or networked sensing architectures~\cite{CarlosAVS2019}.
Recent experiments have explored this possibility using phonon interferometry, where counter-propagating sound modes produce a rotation-dependent precession of an imprinted density pattern~\cite{MartiPRA2015,WoffindenSCIPOST2023,FernándezARXIV2025}.
This technique has recently allowed the measurement of the quantum of circulation~\cite{KumarNJP2016} in fermionic superfluids~\cite{FernándezARXIV2025}.
However, the damping and finite lifetime of the imprinted excitation limit both the interrogation time and the stability of rotation frequency estimation~\cite{MartiPRA2015,WoffindenSCIPOST2023}.

\begin{figure}[b!]
\includegraphics[width=\columnwidth]{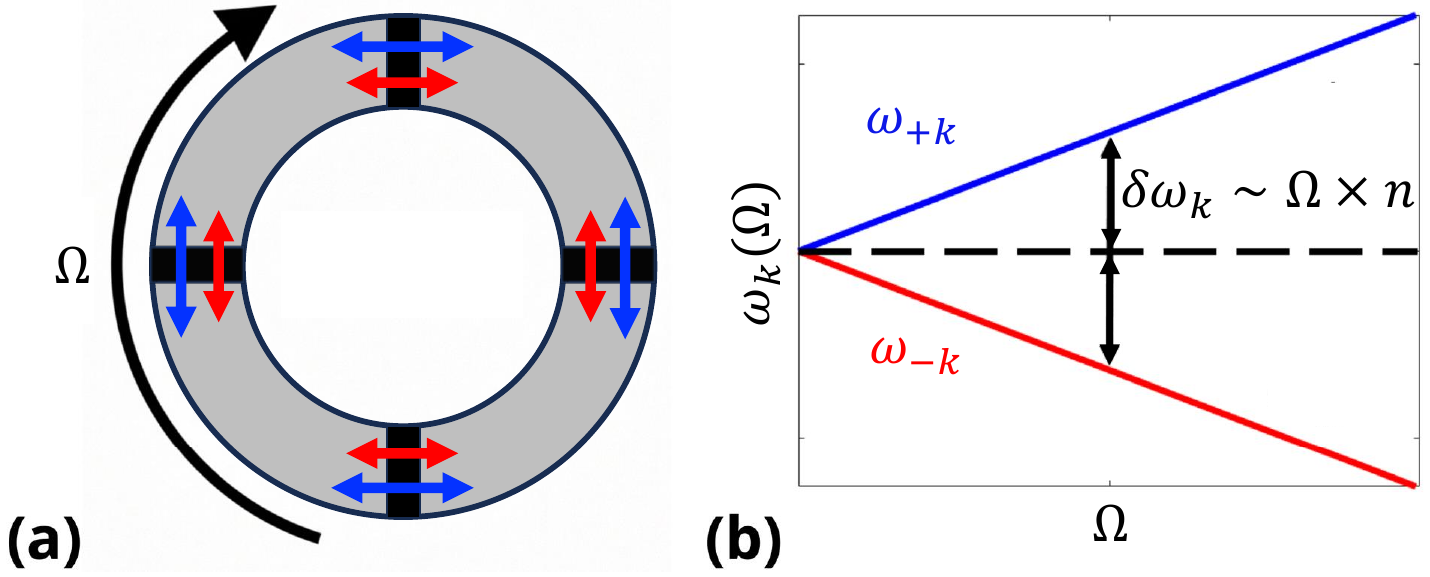}
\caption{
(a) Schematic of a toroidal superfluid interrupted by $n=4$ tunneling barriers 
(black rectangles) and rotating at angular frequency $\Omega$.  
(b) Due to the Doppler shift, a monochromatic (at $\Omega=0$) Josephson mode is split in two frequency modes $\omega_{\pm k}(\Omega)$ [blue and red lines corresponding schematically to blue and red arrows in panel (a)].
The splitting $\omega_{+k}(\Omega)-\omega_{- k}(\Omega) = 2\delta\omega_k(\Omega)$ is proportional to $\Omega$ and the number of 
barriers $n$.
}
\label{Figure1}
\end{figure}

Here we propose a different route to rotation sensing based on the collective Josephson spectrum of a multi-junction toroidal superfluid.
We consider a ring Bose-Einstein condensate (BEC) equipped with $n$ equally spaced tunneling barriers and rotating at angular frequency $\Omega$ with respect to the inertial frame, see Fig.~\ref{Figure1}(a).
In this geometry, rotation lifts the degeneracy of selected small-amplitude Josephson modes.
The resulting Doppler splitting is linear in $\Omega$ and enhanced by the number of junctions $n$, as illustrated in Fig.~\ref{Figure1}(b).
This realizes a spectroscopic form of superfluid rotation sensing: the signal is inferred from the frequency splitting of long-lived population-phase oscillations between weakly linked regions of the ring, rather than from a directly accumulated Sagnac phase or from the precession of a phonon pattern.
We derive analytically the Josephson-mode spectrum and identify population-imbalance patterns that selectively excite the rotation-sensitive modes.
The theory is validated by numerical simulations including the finite transverse width of the toroidal trap, damping, trap imperfections, and quench-induced excitations.
We show that the rotation frequency can be estimated with uncertainty
\begin{equation}
\label{sensitivity_intro}
\Delta\Omega =
\frac{\beta}{(\delta N/\chi) T \sqrt{N_p}, n^{3/2}},
\end{equation}
where $\beta$ is a geometry-dependent factor set by the barrier height and width, $T$ is the observation time, $N_p$ is the number of independent samples, $\delta N$ is the initial population imbalance driving the Josephson dynamics, and $\chi$ is the atom-number detection noise.
Key advantages of the scheme include a large signal-to-noise ratio $\delta N/\chi$, long-lived Josephson oscillations that are resilient to imperfections and weakly coupled to phonon modes, and a favorable $n^{3/2}$ enhancement with the number of sites, where $n\simeq 20$ in current experiments~\cite{PezzeNATCOMM2024}.

{\it Toroidal superfluid.---}   
We consider a BEC in a toroidal trap rotating at angular frequency $\Omega$ around the vertical axis. 
In the frame co-rotating with the trap, the zero-temperature dynamics is governed by the (dimensionless) Gross-Pitaevskii equation (GPE)
\begin{equation}\label{GPE}
i  \dfrac{d \Psi(\vect{r},t)}{dt} = \biggl(H_0+ g|\Psi(\vect{r},t)|^2 + i\Omega\dfrac{d}{d\theta}\biggr)\,\Psi(\vect{r},t),
\end{equation} 
where $H_0 = -\tfrac{1}{2}\nabla^2 + V(\vect{r})$, $\Psi(\vect{r},t)$ is the condensate wavefunction, and $\theta$ is the azimuthal coordinate. The potential \(V(\vect r)\), sketched in Fig.~\ref{Figure1}(a), is a toroidal box trap with inner and outer radii \(R_{\rm in}\) and \(R_{\rm out}\), interrupted by \(n\) equally spaced tunnel barriers of height \(V_0\) and width \(\sigma_0\). 
This realizes a periodic azimuthal lattice of weak links~\cite{PezzeNATCOMM2024, NestiARXIV2025}. 
In Eq.~(\ref{GPE}) energies and times are rescaled in units of $\hbar^2/mR_{in}^2$ and $mR_{in}^2/\hbar$ respectively. 

{\it Analytical model.---}
We describe the dynamics using a many-mode ansatz
\begin{equation} \label{Ansatz}
    \Psi(\vect r,t)
    =
    \sum_{j=1}^n
    \sqrt{N_j(t)}\,e^{i\phi_j(t)}\,\Phi_j(\vect r),
\end{equation}
where \(j=1,\ldots,n\) labels the weakly linked regions of the ring, 
\(N_j(t)\) and \(\phi_j(t)\) are the atom number and phase at site \(j\), 
and \(\Phi_j(\vect r)\) is a localized single-site real wavefunction. 
We assume the orthonormality condition
$\int d^3\vect r\,\Phi_i^*(\vect r)\Phi_j(\vect r)=\delta_{ij}$.
Such functions can be constructed numerically using localized orbitals, for instance Wannier-like functions~\cite{ZhuPRA2015,MarzariPRB1997}. 
The ansatz \eqref{Ansatz} is normalized to the total atom number $N=\sum_{j=1}^n N_j(t)$. 
It assumes that the relevant dynamics is captured by the populations and phases of the $n$ weakly linked regions, while the local wavefunctions  $\Phi_j(\vect r)$ remain time independent.

Replacing Eq.~(\ref{Ansatz}) into Eq.~(\ref{GPE}), projecting onto the localized basis, and separating real and imaginary parts gives coupled equations for $N_j(t)$ and $\phi_j(t)$. 
We focus on small-amplitude oscillations and linearize the equations around the equilibrium configuration $N_0 = N/n$ and $\phi_0$ by setting $N_j(t) = N_0 + \delta N_j(t)$ and $\phi_j(t) = \phi_0 + \delta \phi_j(t)$. 
Keeping only nearest-neighbor contributions, we obtain the following second-order equation for the population imbalance:
\begin{equation}\label{eq:SecondOrder}
{\delta \ddot{\vect{N}}}(t) = \Big(2JUN_0 \vect{L}- J^2 \vect{L}^2 + 2\alpha \Omega\vect{D}\frac{d}{dt} - \alpha^2 \Omega^2\vect{D}^2\Big)\,\delta\vect{N}(t),
\end{equation}
where $\delta \vect{N}(t) = \{\delta N_1(t), ...,\delta N_n(t)\}$. 
An analogous equation holds for the phase fluctuations $\delta\vect{\phi}(t) = \{\phi_1(t), ..., \phi_n(t) \}$ (see Appendix).
The $n\times n$ matrices $\vect{L}$ and $\vect{D}$ have elements $\vect{L}_{ij} = \delta_{i,j+1} + \delta_{i,j-1} - 2\delta_{ij}$, $\vect{D}_{ij} = \delta_{i,j+1} - \delta_{i,j-1}$, with periodic boundary conditions. 
In deriving Eq.~(\ref{eq:SecondOrder}), we kept only nearest neighbor contributions.
The coefficients
\be \label{coeff1}
J = - \int d^3 \vect{r} ~ \Phi_j(\vect{r}) H_0 \Phi_{j+1}(\vect{r}),\,\,\,\, {\rm and} \,\,\,\, U= g \int d^3 \vect{r} ~ \Phi_j(\vect{r})^4,
\ee
describe nearest-neighbor tunneling and on-site interaction, respectively, while
\begin{equation}
\alpha =  \int  d^3\vect{r} ~ \Phi_i(\vect{r}) \frac{\partial}{\partial\theta} \Phi_{i+1}(\vect{r}). 
\label{eq:theta_def}
\end{equation}
quantifies the rotation-induced coupling between neighboring localized modes.
Because Eq.~\eqref{eq:SecondOrder} is translationally invariant, its normal modes are plane waves,
\begin{equation}\label{eq:ans}
    \delta {N}_{k,j}(t) = \delta N\,e^{ik(j-1)}e^{-i
    \omega_kt}
\end{equation}
where $\delta N$ is a small amplitude.
Periodic boundary conditions impose $k = 2\pi m/n$ with $m \in [-(n/2-1),...0,...,n/2]$ for even $n$, and $m \in [-(n-1)/2,...0,...,(n-1)/2]$ for odd $n$.
%
The corresponding eigenfrequencies are
\be \label{eq:eigenfreq_rotating}
\omega_{k}(\Omega)
=
2\alpha\Omega\sin(k)
+ 4J \sqrt{
\sin^2 \frac{k}{2}\Big(\Lambda +\sin^2 \frac{k}{2}\Big)},
\ee
where $\Lambda = N_0 U/(2J)$.
Equation~(\ref{eq:eigenfreq_rotating}) is one of the main results of this work: it shows that the rotation of the system produces a Doppler shift of the Josephson-mode spectrum: for \(k>0\), \(\omega_k(\Omega)\) increases with respect to its value at 
\(\Omega=0\), whereas \(\omega_{-k}(\Omega)\) decreases.   
See Appendix for a full derivation of Eqs.~(\ref{eq:SecondOrder})-(\ref{eq:eigenfreq_rotating}).

At $\Omega=0$, a specific Josephson mode with quasi-momentum $k$ can be excited by preparing the initial population imbalance $\delta \tilde{N}_{k,j}(0) = [\delta {N}_{k,j}(0) + \delta {N}_{-k,j}(0)]/2 = \delta N \cos(k (j-1))$. 
The $k=0$ mode corresponds to a uniform population shift and is excluded by total-number conservation, namely $\tilde{N}_{0,j}(0)= 0$.
The $k=\pi$ (available for even values of $n$~\cite{footnoteSmerzi})
with $\delta \tilde{N}_{\pi,j}(0) = \delta N \cos(\pi (j-1))$ corresponds to a staggered solution -- that is, $\delta\tilde{\vect{N}}(0) = \delta N\,\{1,-1,...,1,-1\}$ -- oscillating at frequency $\omega_\pi(0)$.
All the remaining modes are degenerate in pairs, with $\omega_k(0) = \omega_{-k}(0)$.
Taking the real part of the symmetric combination $\delta \tilde{N}_{k,j}(t) = [\delta N_{k,j}(t) + \delta N_{-k,j}(t)]/2$ gives
\begin{equation}\label{patt}
\delta\tilde{N}_{k,j}(t) = \delta N\cos[k (j-1)]\cos\big[\omega_k(0) t\big], \quad {\rm for}~\,\Omega=0.
\end{equation}
Thus, in the absence of rotation, each real population pattern oscillates 
monochromatically at the Josephson frequency $\omega_k(0)$. 
In particular, the modes $k=\pm\pi/2$, available when $n$ is a multiple of four, have nodes in the population imbalance at alternating sites, namely $\delta \tilde{N}_{\pi/2,j}(t)=0$ for some values of $j$, due to perfectly destructive interference.

\begin{figure*}[t!]
\includegraphics[width=1.9\columnwidth]{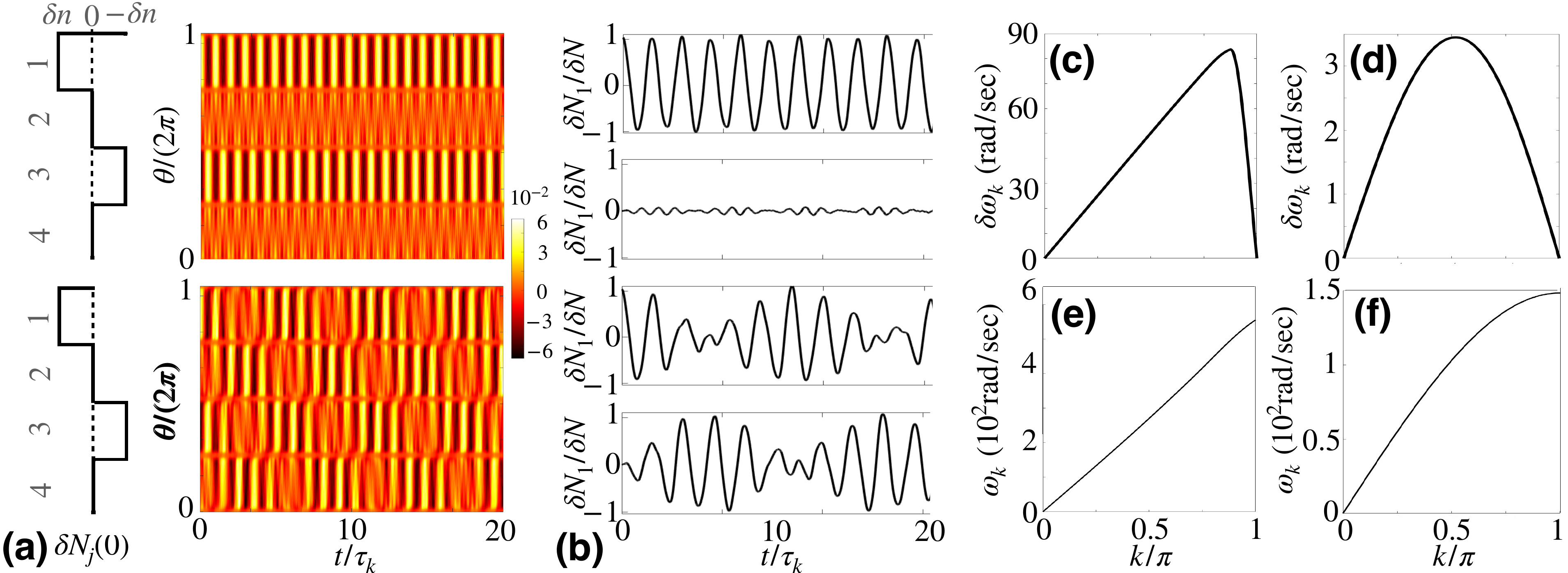}
\caption{
Josephson oscillations from GPE simulations. 
(a) Azimuthal density profile as a function of the angle $\theta$ along the ring and of time, for $\Omega=0$ (top) and $\Omega=2 \pi$ $\rm{rad/sec}$ (bottom). 
The leftmost sketches show the initial population imbalance, $\delta N_{j,k}(0)$ for  $k=\pi/2$. 
The color scale is the rescaled density around the equilibrium value (see Appendix for definition).
(b) Corresponding population oscillations for lattice sites $1$ and $2$, normalized to the initial imbalance $\delta N$. 
In panels (a) and (b), time is rescaled by $\tau_k=2\pi/\omega_k(0)$, with $k=\pi/2$. 
(c,d) Rotation-induced frequency splitting $\delta\omega_k(\Omega)$ as a function of the quasi-momentum $k$.
(e,f) Josephson frequency $\omega_k(0)$ as a function of $k$. 
Left panels, (c) and (e), are obtained for $V_0/\mu=0.3$; right panels, (d) and (f), for $V_0/\mu=6$, where $\mu$ is the chemical potential. 
The parameters used in the GPE simulations are reported in the Appendix.}
\label{Figure2}
\end{figure*}

The situation changes when the same initial population imbalance,
$\delta \tilde N_{k,j}(0)=\delta N\cos[k(j-1)]$, is prepared and the system is allowed to evolve at \(\Omega\neq0\). 
In this case, the Doppler shift lifts the degeneracy between the $\pm k$ modes (except $k=0$ and $\pi$, which are independent from $\Omega$), see the schematic plot in Fig.~\ref{Figure1}(a). 
The population imbalance evolves according to
\begin{equation} \label{oscOmega}
\delta\tilde{N}_{k,j}(t) = \delta N\cos \big[k (j-1)-\delta \omega_k(\Omega) t\big]\cos\big[\omega_k(0) t\big], \,\,\, {\rm for}~\,\Omega \neq 0,
\end{equation}
where $\delta \omega_k(\Omega) = [\omega_{k}(\Omega) - \omega_{-k}(\Omega)]/2 = 2\alpha\Omega \sin k$. 
Notice that $\omega_k(0) = [\omega_{k}(\Omega) + \omega_{-k}(\Omega)]/2$.
Equation~\eqref{oscOmega} shows that rotation transforms the monochromatic Josephson oscillation at \(\Omega=0\) into a two-frequency beating. 
A scaling argument further clarifies the role of the number of barriers, see Appendix. 
For a ring of fixed radius, transverse confinement and total atom number, increasing $n$ reduces the size of each weakly linked region, leading to \(J\sim n^2\), \(U\sim n\), and \(\alpha\sim n\). 
For modes with \(k\) independent of \(n\), such as \(k=\pi/2\), this implies
$\omega_k(0)\sim n$ for $\Lambda\gg1$, $\omega_k(0)\sim n^2$ for $\Lambda\ll 1$,
while the rotation-induced splitting scales as
$\delta\omega_k(\Omega)\sim n\Omega$.

\begin{figure}[t!]
\includegraphics[width=0.95\columnwidth]{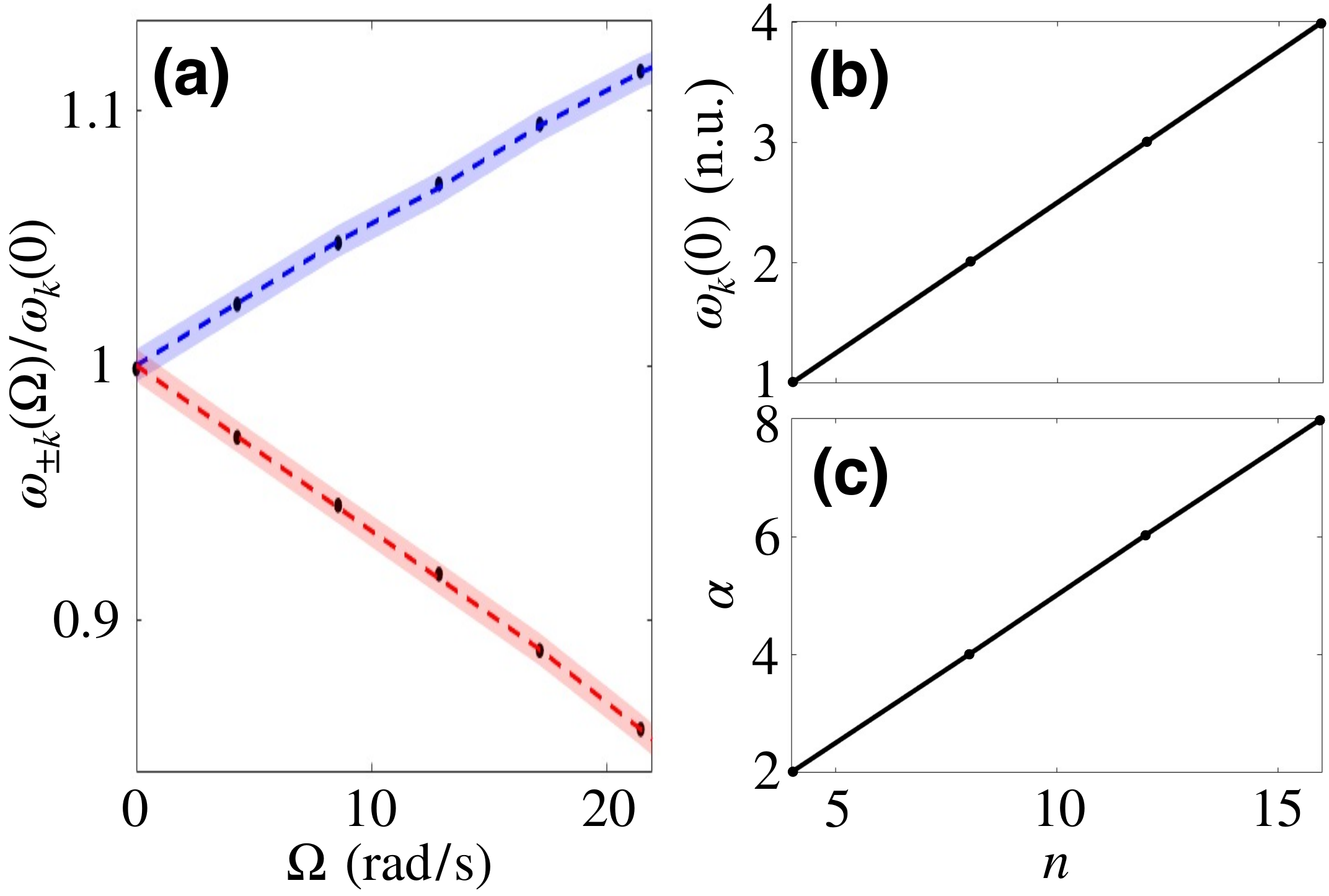}
\caption{
(a) Oscillation frequencies $\omega_{+k}(\Omega)$ (blue dots) and $\omega_{-k}(\Omega)$ (red), normalized to $\omega_k(0)$, as a function of the rotation frequency $\Omega$, for $n=4$. 
Results are obtained from GPE simulations; dashed lines are guides to the eye. 
The shaded regions indicate the Fourier width of the peaks at $\omega_{\pm k}(\Omega)$. 
(b) Josephson frequency $\omega_k(0)$ at $\Omega=0$ as a function of the number of lattice sites $n$. 
Here $\omega_k{(0)}$ is normalized to the value at $n=4$.
(c) Dimensionless coefficient $\alpha=\tfrac12 d\delta\omega_k/d\Omega$ as a function of $n$, extracted at $\Omega=2\pi~{\rm rad/s}$. 
In all panels, the initial population imbalance is chosen to excite the $k=\pi/2$ mode. 
The parameters used in the GPE simulations are reported in the Appendix.
}
\label{Figure3}
\end{figure}

{\it Numerical simulations.---}
We test the predictions of the analytical model with numerical simulations of Eq.~(\ref{GPE}).
Assuming tight harmonic confinement along the (z) axis, Eq.~(\ref{GPE}) reduces to an effective two-dimensional equation.
We prepare a stationary state in the rotating frame with a prescribed initial population imbalance $N_{j,k}(0) = \delta N \cos(k (j-1))$, chosen to excite a target Josephson mode $k$.
To impose this population pattern, we add site-dependent constant energy offsets $V_j$ to the trapping potential $V(\vect r)$, as detailed in the Appendix.
This allows us to control the initial atom number $N_j(0)$ in each lattice site.
At $t=0$, all offsets are switched off, leaving only the lattice potential $V(\vect r)$, and the ensuing Josephson population-phase dynamics is monitored.

In Figs.~\ref{Figure2}(a) and~\ref{Figure2}(b), we initialize the $k=\pi/2$ mode for $n=4$ sites, corresponding to $\delta\vect N(0)=\delta N\{1,0,-1,0\}$.
Figure~\ref{Figure2}(a) shows the integrated azimuthal density profile, while Fig.~\ref{Figure2}(b) shows the corresponding site populations $\delta N_j(t)$, obtained by further integrating the density over each site.
The top panels correspond to $\Omega=0$, while the bottom panels to $\Omega\neq0$.
As predicted by the model, for $\Omega=0$ the populations of the sites with zero initial imbalance, namely sites $2$ and $4$, remain essentially constant, whereas the other sites undergo harmonic Josephson oscillations.
For $\Omega\neq0$, instead, the dynamics displays two dominant frequencies, in agreement with Eq.~(\ref{oscOmega}).
This is confirmed by Fourier analysis, which reveals two main peaks that we identify with $\omega_{\pm k}(\Omega)$.
The two peaks are symmetric around $\omega_k(0)$, and their frequency difference gives $\delta\omega_k(\Omega)$.
We also observe small contributions from other frequencies, mainly due to the initial quench that triggers the dynamics.

Figures~\ref{Figure2}(c) and~\ref{Figure2}(d) report $\delta\omega_k(\Omega)$, while Figs.~\ref{Figure2}(e) and~\ref{Figure2}(f) show $\omega_k(0)$, both as functions of the quasi-momentum $k$, for $n=16$ and fixed $\Omega$.
The panels correspond to two different tunneling-barrier heights: $V_0/\mu=0.3$, (c,e) and $V_0/\mu=6$ (d,f), where $\mu$ is the chemical potential.
In both cases, $\delta\omega_k(\Omega)$ shows a clear dependence on $k$.
For low barriers, $\delta\omega_k(\Omega)$ increases almost linearly with $k$.
The sinusoidal dependence predicted by the analytical model, Eq.~(\ref{eq:eigenfreq_rotating}), is recovered in the tight-binding regime $V_0/\mu\gtrsim1$.
In this regime, $\delta\omega_k(\Omega)$ reaches its maximum at $k=\pi/2$.
Figures~\ref{Figure2}(e) and~\ref{Figure2}(f) show that the reference dispersion $\omega_k(0)$ evolves from an almost linear dispersion for weak barriers to a band-like dispersion for strong barriers.
Overall, Fig.~\ref{Figure2} demonstrates that specific Josephson modes of the GPE can be selectively excited by controlling the initial population imbalance, enabling spectroscopy of the Josephson frequencies both at $\Omega=0$ as well as at finite rotation $\Omega \neq 0$.

In Fig.~\ref{Figure3}(a), we plot $\delta\omega_k(\Omega)$ as a function of $\Omega$.
In these simulations, we prepare the $k=\pi/2$ mode and follow the dynamics up to an observation time $T$ long enough to resolve several beatings.
The frequencies are extracted from Fourier analysis, whose finite resolution is of order $1/T$ (shaded regions).
As predicted by the analytical model, $\delta\omega_k(\Omega)$ increases linearly with $\Omega$.
This allows us to extract $\alpha=\tfrac12 d\delta\omega_k/d\Omega$.
We find that the slope $\alpha$ obtained from the Gross--Pitaevskii simulations increases with the number of sites $n$, consistently with Eq.~(\ref{eq:theta_def}); the results are shown in Fig.~\ref{Figure3}(c).
We also find that $\alpha$ is approximately independent of the interaction strength and of the system size, namely the radius and width of the toroidal trap.
Its dependence on the barrier height is weak, although $\alpha$ decreases for $V_0/\mu\gg1$.
Finally, Fig.~\ref{Figure3}(b) shows that $\omega_k(0)$ increases approximately linearly with $n$, consistently with the scaling argument discussed above, in the regime $\Lambda\gg 1$.
A quadratic bending is observed for much larger values of $n$ (not shown), when approaching $\Lambda\ll 1$.

\begin{figure}[t!]
\includegraphics[width=\columnwidth]{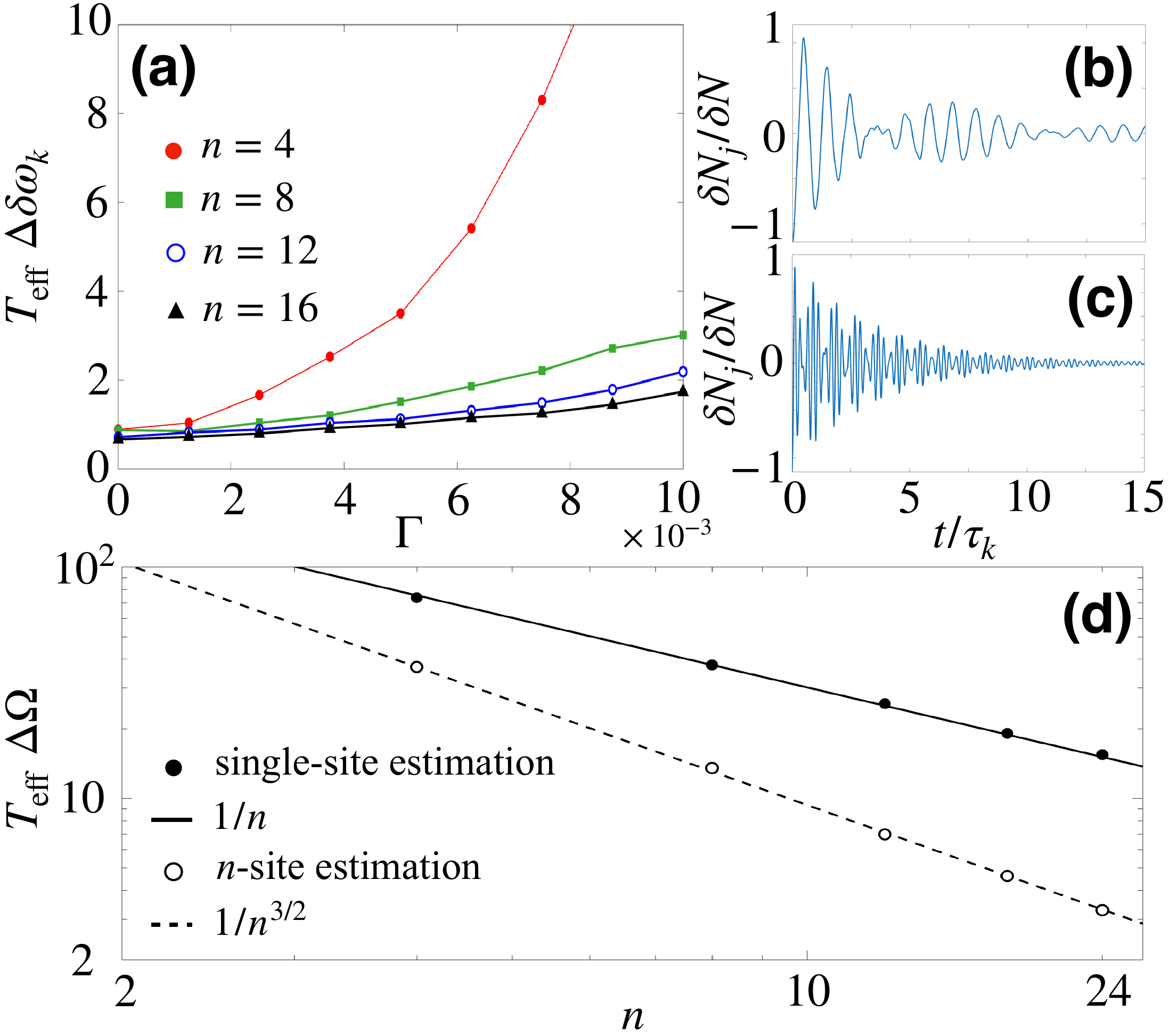}
\caption{Sensitivity to rotations obtained from fits of GPE oscillations.
(a) Fit uncertainty $\Delta\delta\omega_k$ as a function of the dimensionless damping parameter $\Gamma$, defined in the Appendix. 
The comparison between the different $n$ is done at fixed $T_{\rm eff}$.
Different symbols correspond to different numbers of lattice sites $n$; solid lines are guides to the eye.
(b,c) Damped population oscillations for $n=4$ (b) and $n=16$ (c), computed for $\Gamma=5 \times 10^{-3}$. 
In both panels, time is rescaled by $\tau_k=2\pi/\omega_k(0)$, with $\omega_k(0)$ evaluated for the $n=4$ case. 
Although the damping rate increases with $n$, the Josephson frequency also increases, allowing for a more accurate determination of $\delta\omega_k$ from the fit, as shown in panel (a).
(d) Uncertainty $\Delta \Omega$ as a function of $n$. 
Filled dots are obtained by fitting the oscillations of a single site, giving the scaling $T_{\rm eff} \Delta \Omega \sim 1/n$; open circles are obtained by combining the synchronous signals from all sites, giving $T_{\rm eff} \Delta \Omega \sim 1/n^{3/2}$. 
Solid and dashed lines show the corresponding scaling laws. 
Here $\Omega = 2$ rad/s and the parameters used in the simulations are reported in the Appendix.
}
\label{Figure4}
\end{figure}

{\it Sensitivity to rotations.---} 
The dependence of $\delta \omega_k(\Omega)$ on $\Omega$ can be exploited for rotation sensing.
The estimation sequence is as follows: 
(i) prepare an initial population imbalance $\delta N_j(0)=\delta N\cos[k(j-1)]$, restricting in the following to $k=\pi/2$; 
(ii) record the oscillations $\delta N_j(t)$ at each site, with $j=1,\ldots,n$, at $N_p$ intervals, up to an observation time $T$; 
and (iii) extract the oscillation frequencies, for instance by fitting the data with the two-frequency form of Eq.~(\ref{oscOmega}).
To model detection noise and imperfections in a possible experimental sequence, we add Gaussian white noise with amplitude $\chi$ to the Gross--Pitaevskii values of $\delta N_j(t)$.
The sensitivity to $\Omega$ is then obtained from error propagation,
\be 
\label{sensitivity}
\Delta \Omega_j 
= 
\frac{\Delta(\delta \omega_k)}{d(\delta \omega_k)/d\Omega} 
= 
\frac{\beta}{T_{\rm eff} n}, \quad {\rm with}\,\, T_{\rm eff} = \frac{\delta N}{\chi}\sqrt{N_p}T,
\ee
where the second equality uses the standard uncertainty of a frequency fit and is the same for each lattice site.
The scaling $1/n$ arises from the contribution of $d(\delta \omega_k)/d\Omega$.

We simulate the estimation protocol by sampling oscillations obtained from a damped GPE with dimensionaless damping rate $\Gamma$.
The population oscillations decay in time, with a rate that increases with $\Gamma$ (see Appendix).
Consequently, damping limits the maximum useful interrogation time $T$ over which Josephson oscillations can be resolved.
In Fig.~\ref{Figure4}(a) we report the uncertainty of the fit, $\Delta \delta \omega_k$ as a function of $\Gamma$, where different curves are obtained for different number of sites $n$.
As expected, the $\Delta \delta \omega_k$ spreads  with damping.
However, this increase becomes weaker for larger $n$, showing that Josephson necklaces with more barriers are more robust against damping.
This behavior follows from the increase of the Josephson frequency with $n$, shown in Fig.~\ref{Figure4}(b), obtained for $n=4$, and (c), obtained for $n=16$.
For a fixed observation window $T$, larger values of $n$ allow more oscillation cycles and beatings to be observed, improving the frequency resolution of the fit and therefore reducing the uncertainty in the determination of $\delta \omega_k$ from the fit.

Increasing $n$ provides a second advantage: the Josephson dynamics can be monitored simultaneously in all $n$ sites, since a single toroidal condensate provides $n$ synchronous oscillating signals.
Fitting all site populations gives an additional statistical gain, $\Delta \Omega \sim \Delta \Omega_j/\sqrt{n}$, provided that the detection noise is uncorrelated between sites.
Therefore, combining this result with Eq.~(\ref{sensitivity}) gives Eq.~(\ref{sensitivity_intro}).
Figure~\ref{Figure4}(d) confirms numerically this prediction. 
We plot $\Delta \Omega$ as a function of $n$, obtained from a numerical analysis of a single (circles) and $n$ populations (dots) of a single toroidal system with $n$ lattice sites.
The multi-site analysis confirms the predicted $1/n^{3/2}$ scaling. 

Let us finally give a concrete value for $\Delta \Omega$. 
Form the results of Fig.~(\ref{Figure4}) we find $\beta \approx 250$. 
Taking $N=2\times 10^6$ total atoms and $n=20$ sites, and assuming that we can excite a Josephson mode with a contrast $\delta N/N_0 =0.10$, we obtain $\delta N/\chi = 0.1\sqrt{N_0}\approx 32$ for shot-noise limited detection.
Considering $N_p=50$ measurements and $T=400$ msec, we find $\Delta \Omega = 0.02$ rad/sec, which is more than one order of magnitude below the sensitivity reported experimentally in Refs.~\cite{MartiPRA2015, WoffindenSCIPOST2023}. 
By further optimizing the system (barrier height, width, interaction strength, geometry, etc) we expect to further reduce $\Delta \Omega$. 

{\it Conclusions.---}
To summarize, we predict the normal modes of  Josephson oscillations in a ring superfluid and discuss a method to observe the Doppler splitting in the rotating system. 
The latter effect, increasing with $n$ and $\Omega$, can be exploited to realize a  micrometer-size gyroscope. 
The system can be understood as a ``discretized'' version of the phonon interferometry of Refs.~\cite{MartiPRA2015, WoffindenSCIPOST2023, KumarNJP2016, FernándezARXIV2025}.
In our case, the dynamics is controlled by tunneling through weak links rather than by unconstrained nonlinear propagation, and it is thus expected to be more resilient to excitations.

Our predictions can be readily tested experimentally  at various levels.
At $\Omega=0$, we predict interesting interference effects and scalings of Josephson modes that would extend the study of population-phase oscillations~\cite{CataliottiSCIENCE2001} from double-well~\cite{AlbiezPRL2005, ValtolinaSCIENCE2015, SpagnolliPRL2017} to a toroidal system.
Proof-of-principle predictions at $\Omega\neq 0$ can be tested with rotating barriers, which can be implemented with a digital micromirror device~\cite{DelPaceSCIENCE2025}. 
Moreover, thanks to the close analogy with bosonic Josephson lattices~\cite{BiagioniNATURE2024, DonelliPRA2025}, our analysis can be extended to dipolar supersolids in ring traps~\cite{TengstrandPRA2021, TengstrandPRA2023, ŠindikPRL2024, PretiPRL2026}, as well as exciton-polariton rings systems~\cite{VoronovaNATCOMM2025}.
Finally, building on our mean-field analysis, it would be possible to include quantum fluctuations following approaches previously developed for the double-well case~\cite{GatiJPB2007} and eventually explore sensitivity enhancement due to squeezing and entanglement~\cite{EstèveNATURE2008, BerradaNATCOMM2013, PezzeRMP2018}.


{\it Acknowledgments.---}We acknowledge support from the Horizon Europe programme HORIZON-CL4-2022-QUANTUM-02-SGA via the project 10113690 PASQuanS2.1.

\section{Appendix}

{\it Linear model.---} 
Substituting the many-mode expansion Eq.~\eqref{Ansatz} into the Gross--Pitaevskii equation \eqref{GPE}, and separating real and imaginary parts, gives
\begin{align}
\dot{N}_i &= \sum_{j=\pm1} \Bigl\{ -2J\sqrt{N_i N_{i+j}}\sin(\phi_{i+j}-\phi_i) \nonumber \\
&\qquad\quad + 2j\alpha\Omega\sqrt{N_i N_{i+j}}\cos(\phi_{i+j}-\phi_i) \Bigr\}, \label{eq:nonlinear_N} \\
\dot{\phi}_i &= -E_0 - N_i U + \sum_{j=\pm1} \Bigl\{ J\sqrt{\tfrac{N_{i+j}}{N_i}}\cos(\phi_{i+j}-\phi_i) \nonumber \\
&\qquad\quad + j\alpha\Omega\sqrt{\tfrac{N_{i+j}}{N_i}}\sin(\phi_{i+j}-\phi_i) \Bigr\}, \label{eq:nonlinear_phi}
\end{align}
where
\begin{equation}
E_0 =
\int d^3 r\,
\Phi_i^*(\mathbf r) H_0 \Phi_i(\mathbf r)
\end{equation}
is the on-site single-particle energy, independent of index \(i\) for a
translationally invariant ring.
This is a nonlinear system for the site populations \(N_i\) and phases \(\phi_i\). 
We linearize Eqs.~\eqref{eq:nonlinear_N} and~\eqref{eq:nonlinear_phi} around the uniform equilibrium state by writing \(N_i=N_0+\delta N_i\) and \(\phi_i=\phi_0+\delta\phi_i\), with small population and phase fluctuations. 
After expanding to first order and removing the uniform global phase evolution, we obtain
\begin{align}
\dot{\delta N}_i(t) &= -2N_0 J(\delta\phi_{i+1} + \delta\phi_{i-1} - 2\delta\phi_i) \nonumber \\
&\quad + \alpha\Omega(\delta N_{i+1} - \delta N_{i-1}), \label{eq:linear_N} \\
\dot{\delta \phi}_i(t) &= -U\delta N_i + \frac{J}{2N_0}(\delta N_{i+1} + \delta N_{i-1} - 2\delta N_i) \nonumber \\
&\quad + \alpha\Omega(\delta\phi_{i+1} - \delta\phi_{i-1}). \label{eq:linear_phi}
\end{align}
Introducing the discrete Laplacian \(({L}\vect{x})_i=x_{i+1}+x_{i-1}-2x_i\) and the discrete derivative \(({D}\vect{x})_i=x_{i+1}-x_{i-1}\), the linear system can be written in compact form as
\begin{equation} 
\label{eq:matrix_rotating}
\begin{pmatrix}
\dot{\delta\vect{N}}(t) \\
\dot{\delta\vect{\phi}}(t)
\end{pmatrix}
=
\begin{pmatrix}
\alpha \Omega \vect{D} & -2N_0 J \vect{L} \\
-U \Eins + \frac{J}{2N_0} \vect{L} & \alpha \Omega \vect{D}
\end{pmatrix}
\begin{pmatrix}
\delta\vect{N}(t) \\
\delta\vect{\phi}(t)
\end{pmatrix},
\end{equation}
where \(\Eins\) is the \(n\times n\) identity matrix, 
\(\delta\vect{N}(t)=\{\delta N_1(t),\ldots,\delta N_n(t)\}\), and 
\(\delta\vect{\phi}(t)=\{\delta \phi_1(t),\ldots,\delta \phi_n(t)\}\).
We now derive Eq.~\eqref{eq:SecondOrder}. 
Differentiating Eq.~\eqref{eq:linear_N} with respect to time gives
\begin{equation}
\ddot{\delta N}_i(t) = -2N_0 J \frac{d}{dt}\big(\delta \phi_{i+1} + \delta \phi_{i-1} - 2\delta \phi_i\big) + \alpha\Omega \frac{d}{dt}\big(\delta N_{i+1} - \delta N_{i-1}\big).
\label{eq:ddN_start}
\end{equation}
The time derivative of the phase fluctuation is evaluated using Eq.~\eqref{eq:linear_phi}. 
Substituting this expression into the first term of Eq.~\eqref{eq:ddN_start} yields
\begin{align}
-2N_0 J \big(\dot{\delta \phi}_{i+1} &+ \dot{\delta \phi}_{i-1} - 2\dot{\delta \phi}_i\big) = 
2JUN_0(\delta N_{i+1} + \delta N_{i-1} - 2\delta N_i) \nonumber \\
&\quad - J^2\big(\delta N_{i+2} + \delta N_{i-2} - 4\delta N_{i+1} - 4\delta N_{i-1} + 6\delta N_i\big) \nonumber \\
&\quad - 2N_0 J \alpha \Omega\big(\delta \phi_{i+2} - \delta \phi_{i-2} - 2\delta \phi_{i+1} + 2\delta \phi_{i-1}\big).
\label{eq:sub_dphi}
\end{align}
The last term in Eq.~\eqref{eq:sub_dphi} is eliminated using Eq.~\eqref{eq:linear_N}, giving
\begin{equation}
\begin{split}
\label{secondneigh}
-2N_0 J \alpha\Omega\big(&\delta \phi_{i+2} - \delta \phi_{i-2} - 2\delta \phi_{i+1} + 2\delta \phi_{i-1}\big) = \\
&\alpha\Omega\big(\dot{\delta N}_{i+1} - \dot{\delta N}_{i-1}\big) - \alpha^2\Omega^2\big(\delta N_{i+2} + \delta N_{i-2} - 2\delta N_i\big).
\end{split}
\end{equation}
Combining Eqs.~\eqref{eq:ddN_start}, \eqref{eq:sub_dphi}, and \eqref{secondneigh}, we obtain the second-order equation for the population fluctuations:
\begin{align}
\ddot{\delta N}_i(t) &= 2JUN_0(\delta N_{i+1} + \delta N_{i-1} - 2\delta N_i) \nonumber \\
&\quad - J^2\big(\delta N_{i+2} + \delta N_{i-2} - 4\delta N_{i+1} - 4\delta N_{i-1} + 6\delta N_i\big) \nonumber \\
&\quad + 2\alpha\Omega\big(\dot{\delta N}_{i+1} - \dot{\delta N}_{i-1}\big) - \alpha^2\Omega^2\big(\delta N_{i+2} - 2\delta N_i + \delta N_{i-2}\big).
\label{eq:ddN_explicit}
\end{align}
Recognizing the action of \(\vect{L}\), \(\vect{L}^2\), \(\vect{D}\), and \(\vect{D}^2\) in Eq.~\eqref{eq:ddN_explicit} gives Eq.~\eqref{eq:SecondOrder} of the main text.

The analogous calculation for the phase fluctuations is obtained by differentiating Eq.~\eqref{eq:linear_phi} and substituting Eq.~\eqref{eq:linear_N}. 
This gives
\begin{equation}
\label{eq:SecondOrderphi}
{\delta \ddot{\vect{\phi}}}(t) = \Big(2JUN_0 \vect{L}- J^2 \vect{L}^2 + 2\alpha \Omega\vect{D}\frac{d}{dt} - \alpha^2 \Omega^2\vect{D}^2\Big)\,\delta\vect{\phi}(t).
\end{equation}
Thus, both population and phase fluctuations obey the same second-order linear operator and admit plane-wave normal-mode solutions.

Finally, we derive the eigenfrequencies. 
Inserting the plane-wave ansatz Eq.~\eqref{eq:ans} into Eq.~\eqref{eq:ddN_explicit} gives
\begin{equation}
\label{omegsq}
-\omega_k^2
=
-2JUN_0q_k
-J^2q_k^2
+4\alpha^2\Omega^2\sin^2k
-4\alpha\Omega\,\omega_k\sin k,
\end{equation}
where $q_k = 4\sin^2\left(\frac{k}{2}\right)$.
Solving Eq.~\eqref{omegsq} and retaining the positive-frequency branch gives Eq.~\eqref{eq:eigenfreq_rotating} of the main text. \\


{\it Scaling of coefficients.---}
The scaling with the number of barriers can be understood from a simple one-dimensional argument. 
Consider a ring coordinate \(\theta\in(0,2\pi)\), and assume that each localized orbital is essentially supported within a single well. 
The normalization condition is then
\begin{equation}
\int_0^{2\pi} d\theta\,|\Phi_j(\theta)|^2
\simeq
\int_{\mathrm{well}\,j} d\theta\,|\Phi_j(\theta)|^2
=1 .
\end{equation}
We introduce the rescaled coordinate
$\theta=\theta_j+\frac{2\pi}{n}x$, with $x\in(-1/2, 1/2)$, where \(\theta_j\) is the center of the \(j\)-th well. 
Assuming that the localized mode has an \(n\)-independent shape in the rescaled coordinate, $\Phi_j(\theta)=A_n\,\tilde\phi(x)$, we obtain
\begin{equation}
1
=
A_n^2\frac{2\pi}{n}
\int_{-1/2}^{1/2} dx\,|\tilde\phi(x)|^2 .
\end{equation}
Since the integral is independent of \(n\), this gives $A_n\propto \sqrt n$.

This immediately gives the scaling of the on-site interaction:
\begin{equation}
U
\propto
\int d\theta\,|\Phi_j(\theta)|^4
=
A_n^4\frac{2\pi}{n}
\int_{-1/2}^{1/2} dx\,|\tilde\phi(x)|^4
\propto n .
\end{equation}
Similarly, the rotation-induced coefficient scales as
\begin{equation}
\alpha
\propto
\int d\theta\,
\Phi_j(\theta)
\frac{\partial}{\partial\theta}
\Phi_{j+1}(\theta).
\end{equation}
Using \(\partial_\theta=(n/2\pi)\partial_x\), one finds
\begin{equation}
\alpha
\propto
A_n^2
\frac{n}{2\pi}
\frac{2\pi}{n}
\int dx\,\tilde\phi_j(x)\partial_x\tilde\phi_{j+1}(x)
\propto A_n^2
\propto n .
\end{equation}
Finally, the tunneling coefficient \(J\) is dominated by the kinetic-energy contribution. 
Since
\begin{equation}
\frac{\partial^2}{\partial\theta^2}
=
\left(\frac{n}{2\pi}\right)^2
\frac{\partial^2}{\partial x^2},
\end{equation}
we have
\begin{equation}
J
\propto
\int d\theta\,
\Phi_j(\theta)
\left(-\frac{\partial^2}{\partial\theta^2}\right)
\Phi_{j+1}(\theta)
\propto
A_n^2
\left(\frac{n}{2\pi}\right)^2
\frac{2\pi}{n}
\propto n^2 .
\end{equation}

{\it Simulations details.---}
The external potential $V(\theta, r)$ consists of $n$ square barriers added to a background offset. Barrier centers are located at
\begin{equation}
\theta_j = \frac{2\pi j}{N_B} + \phi_0, \quad j = 0, \dots, N_B-1,
\end{equation}
If the angular half-width of each barrier has a value of $\omega_\theta$, the total potential can be defined as
\begin{equation}\label{pot}
V(\theta, r) = 
\begin{cases}
V_0 & \text{for } \theta_j - \omega_\theta<\theta<\theta_j+\omega_\theta\\
V^{\text{well} }_{j}& \text{if } \theta \in \text{well } j \text{ (outside barriers)},
\end{cases}
\end{equation}
where $V^{\text{well} }_{j}$ is the static offset applied to the $j$-th well region.
Equation~(\ref{GPE}) is first solved with the full potential Eq.~(\ref{pot}). 
Oscillatory dynamics is then triggered by putting $V^{\text{well} }_{j} = 0$ for all $j$ and evolving the wavefunction in real time.

The main equation which is numerically addressed in this paper is the $2$-D GPE
\begin{equation}\label{2DGPE2}
\left( -\frac{\hbar^2}{2m}\nabla_{2D}^2 + V(r,\theta) + g_{2D}|\psi_{2D}|^2 \right)\psi_{2D} = i\hbar\frac{\partial \psi_{2D}}{\partial t}
\end{equation}
obtained from Eq.~(\ref{GPE}) by introducing a strong harmonic confinement along the $z$-direction. The parameter $g_{2D} = \sqrt{8\pi}{\hbar^2 a_s}/{m a_z}$ is the effective $2$-D interaction constant, with $a_z = \sqrt{\hbar/m\omega_z}$. 
The $1$-D density profiles can be obtained as $\rho(\theta,t) = |\Psi(\theta,t)|^2 = \int rdr\rho(r,\theta,t)$, where $\rho(r,\theta,t) = |\Psi_{2D}(r,\theta,t)|^2$. 

Figure~\ref{Figure2}(a) shows two $1$-D density profiles in time for a $k = \pi/2$ configuration with $n = 4$, that is, having $\delta \vect{N} (0) = \delta n\cdot(1,0,-1,0)$.
The data are normalized to the average spatial value $\bar{\rho}$ and shifted by the time average, $\tilde{\rho}(\theta,t) = \rho(\theta,t)/\bar{\rho}-\braket{\rho(\theta,t)/\bar{\rho}}_t$. Top (bottom) panel is obtained for $\Omega = 0$ ($\Omega = 2\pi$) rad/s.

Figure~\ref{Figure2}(b) shows number imbalance $\delta N = N_j - N_0$ ($j = 1,2$) as a function of time. Top (bottom) panels are obtained for $\Omega = 0$ ($\Omega = 2\pi$) rad/s. Data is obtained from simulations of Eq.~(\ref{2DGPE2}). Parameters are $N = 4.8\cdot10^4$, $R_{in} = 20\, \mu m$, $R_{out} = 25\, \mu m$, $\omega_z = 3.14\cdot10^3$, $a_s = 5 \,a_0$, $V_0/\mu \sim 1.5$, $\sigma/\xi\sim 1.5$, where $a_0$ is the Bohr radius, $\mu$ is the system's chemical potential while $\xi$ is the bulk healing length.  In dimensionless energy units ($\hbar^2/mR_{in}^2$) the potential bias in each well is set as $V^{well} = (4, 2,0,2)$.

In Fig.~\ref{Figure3}, 
parameters of GPE simulations are set as $N = 3.2\cdot10^4$, $\omega_z = 3.14\cdot 10^3$ rad/s, $a_s = 100\,a_0$, $V_0/\mu \sim 1.5$, $\sigma/\xi\sim 1.25$, $R_{in} = 25\, \mu m$,  $R_{out} = 30\, \mu m$. In dimensionless energy units ($\hbar^2/mR_{in}^2$) the potential bias in each well is set as $V^{well} = (5, 2.5,0,2.5)$.
The total time of simulation is $T = 1$s.

The simulations shown in Fig.~\ref{Figure4}(a) include the effect of a dimensionless phenomenological damping parameter $\Gamma$.
The standard damped GPE is written as
\begin{equation}
    i\hbar\dfrac{\partial\Psi}{\partial t} = (1-i\Gamma)(H_{GPE}-\mu)\Psi
\end{equation}
where $H_{GPE} = (-\hbar^2\nabla^2/2m + V+g|\Psi|^2)$.
In Fig.~\ref{Figure4}, the parameters of GPE simulations are set as $N = 2\cdot10^4$, $\omega_z = 3.14\cdot 10^3$ rad/s, $a_s = 5\,a_0$, $V_0/\mu \sim 0.8$, $\sigma/\xi\sim 0.7$, $R_{in} = 25\, \mu m$,  $R_{out} = 30\, \mu m$. The potential bias in each well is set as $V^{well} = (1, 0.5,0,0.5)$, iterated $n/4$ times. Simulation time is $T = 4$s.

We have checked the scaling of the fit uncertainty $\Delta \delta \omega_k$ with $T$, $\chi/\delta N$ and $\sqrt{N_p}$. 
To facilitate the comparison between different cases in Fig.~\ref{Figure4}(a), we have kept $T_{\rm eff}$ fixed.
In the results presented in Fig.~\ref{Figure4}(b) we did not include damping.

%


\begin{thebibliography}{100}

\bibitem{PoloPR2025}
L.~A.~J. Polo, W.~J. Chetcuti, T. Haug, A. Minguzzi, and K. Wright,
Persistent currents in ultracold gases,
\textit{Phys. Rep.} \textbf{1137}, 1--70 (2025).

\bibitem{RyuPRL2007}
C. Ryu, M.~F. Andersen, P. Cladé, V. Natarajan, K. Helmerson, and W.~D. Phillips,
Observation of persistent flow of a Bose--Einstein condensate in a toroidal trap,
\textit{Phys. Rev. Lett.} \textbf{99}, 260401 (2007).

\bibitem{KumarNJP2016}
A. Kumar, N. Anderson, W.~D. Phillips, S. Eckel, G.~K. Campbell, and S. Stringari,
Minimally destructive, Doppler measurement of a quantized flow in a ring-shaped Bose--Einstein condensate,
\textit{New J. Phys.} \textbf{18}, 025001 (2016).

\bibitem{KumarPRA2018}
A. Kumar, R. Dubessy, T. Badr, C. De Rossi, M. de Goër de Herve, L. Longchambon, and H. Perrin,
Producing superfluid circulation states using phase imprinting,
\textit{Phys. Rev. A} \textbf{97}, 043615 (2018).

\bibitem{CaiPRL2022}
Y. Cai, D.~G. Allman, P. Sabharwal, and K.~C. Wright,
Persistent currents in rings of ultracold fermionic atoms,
\textit{Phys. Rev. Lett.} \textbf{128}, 150401 (2022).

\bibitem{DelPacePRX2022}
G. Del Pace, K. Xhani, A. Muzi Falconi, M. Fedrizzi, N. Grani, D. Hernández-Rajkov, M. Inguscio, F. Scazza, W.~J. Kwon, and G. Roati,
Imprinting persistent currents in tunable fermionic rings,
\textit{Phys. Rev. X} \textbf{12}, 041037 (2022).

\bibitem{WrightPRL2013}
K.~C. Wright, R.~B. Blakestad, C.~J. Lobb, W.~D. Phillips, and G.~K. Campbell,
Driving phase slips in a superfluid atom circuit with a rotating weak link,
\textit{Phys. Rev. Lett.} \textbf{110}, 025302 (2013).

\bibitem{RyuPhys2013}
C. Ryu, P.~W. Blackburn, A.~A. Blinova, and M.~G. Boshier,
Experimental realization of Josephson junctions for an atom SQUID,
\textit{Phys. Rev. Lett.} \textbf{111}, 205301 (2013).

\bibitem{RyuNature2020}
C. Ryu, E.~C. Samson, and M.~G. Boshier,
Quantum interference of currents in an atomtronic SQUID,
\textit{Nat. Commun.} \textbf{11}, 3338 (2020).

\bibitem{RamanathanPhys2011}
A. Ramanathan \textit{et al.},
Superflow in a toroidal Bose--Einstein condensate: An atom circuit with a tunable weak link,
\textit{Phys. Rev. Lett.} \textbf{106}, 130401 (2011).

\bibitem{JendrzejewskiPRL2014}
F. Jendrzejewski, S. Eckel, N. Murray, C. Lanier, M. Edwards, C.~J. Lobb, and G.~K. Campbell,
Resistive flow in a weakly interacting Bose--Einstein condensate,
\textit{Phys. Rev. Lett.} \textbf{113}, 045305 (2014).

\bibitem{EckelPRX2014}
S. Eckel, F. Jendrzejewski, A. Kumar, C.~J. Lobb, and G.~K. Campbell,
Interferometric measurement of the current-phase relationship of a superfluid weak link,
\textit{Phys. Rev. X} \textbf{4}, 031052 (2014).

\bibitem{PezzeNATCOMM2024}
L. Pezz\`e, K. Xhani, C. Daix, N. Grani, B. Donelli, F. Scazza, D. Hernández-Rajkov, W.~J. Kwon, G. Del Pace, and G. Roati,
Stabilizing persistent currents in an atomtronic Josephson junction necklace,
\textit{Nat. Commun.} \textbf{15}, 4831 (2024).

\bibitem{XhaniImpurities2025}
K. Xhani, G. Del Pace, N. Grani, D. Hernández-Rajkov, B. Donelli, G. Roati, and L. Pezz\`e,
Tuning the critical current in toroidal superfluids via controllable impurities,
\textit{Physical Review A} {\bf 113}, L051301 (2026).

\bibitem{GanPRR2026}
K.~S. Gan, V.~P. Singh, L. Amico, and R. Dumke,
Josephson dynamics in two-dimensional ring-shaped condensates,
\textit{Phys. Rev. Research} \textbf{8}, 023190 (2026).

\bibitem{TüzemenPRR2026}
B. Tüzemen, A. Barresi, G. Wlazłowski, P. Magierski, and K. Xhani,
Impurity-controlled vortex mobility and pair breaking in fermionic superfluid rings,
\textit{Phys. Rev. Research} \textbf{8}, 023204 (2026).

\bibitem{BorysenkoPRA2025}
Y. Borysenko, N. Bazhan, O. Prykhodko, D. Pfeiffer, L. Lind, G. Birkl, and A. Yakimenko,
Acceleration-driven dynamics of Josephson vortices in coplanar superfluid rings,
\textit{Phys. Rev. A} \textbf{111}, 043308 (2025).

\bibitem{Hernández-RajkovNATPHYS2024}
D. Hernández-Rajkov, N. Grani, F. Scazza, G. Del Pace, W.~J. Kwon, M. Inguscio, K. Xhani, C. Fort, M. Modugno, F. Marino, and G. Roati,
Connecting shear flow and vortex array instabilities in annular atomic superfluids,
\textit{Nat. Phys.} \textbf{20}, 939--944 (2024).

\bibitem{ChenPRR2025}
K.-J. Chen, W. Yi, and F. Wu,
Dynamic generation of superflow in a fermionic ring through phase imprinting,
\textit{Phys. Rev. Research} \textbf{7}, 013022 (2025).

\bibitem{CiszakARXIV2025}
M. Ciszak, N. Grani, D. Hernández-Rajkov, G. Del Pace, G. Roati, and F. Marino,
Cooperative stabilization of persistent currents in superfluid ring networks,
\textit{arXiv}:2601.15121 (2026).

\bibitem{NestiARXIV2025}
G. Nesti and L. Pezz\`e,
Increasing the stability of a superfluid in a rotating necklace potential,
\textit{Phys. Rev. A} \textbf{113}, 063326 (2026).

\bibitem{MartiPRA2015}
G.~E. Marti, R. Olf, and D.~M. Stamper-Kurn,
Collective excitation interferometry with a toroidal Bose--Einstein condensate,
\textit{Phys. Rev. A} \textbf{91}, 013602 (2015).

\bibitem{WoffindenSCIPOST2023}
C.~W. Woffinden, A.~J. Groszek, G. Gauthier, B.~J. Mommers, M.~W.~J. Bromley, S.~A. Haine, H. Rubinsztein-Dunlop, M.~J. Davis, T.~W. Neely, and M. Baker,
Viability of rotation sensing using phonon interferometry in Bose--Einstein condensates,
\textit{SciPost Phys.} \textbf{15}, 128 (2023).

\bibitem{PackardPRB1992}
R.~E. Packard and S. Vitale,
Principles of superfluid-helium gyroscopes,
\textit{Phys. Rev. B} \textbf{46}, 3540 (1992).

\bibitem{SchwabNATURE1997}
K. Schwab, N. Bruckner, and R.~E. Packard,
Detection of the Earth's rotation using superfluid phase coherence,
\textit{Nature} \textbf{386}, 585 (1997).

\bibitem{SatoRPP2012}
Y. Sato and R.~E. Packard,
Superfluid helium quantum interference devices: Physics and applications,
\textit{Rep. Prog. Phys.} \textbf{75}, 016401 (2012).

\bibitem{GustavsonPRL1997}
T.~L. Gustavson, P. Bouyer, and M.~A. Kasevich,
Precision rotation measurements with an atom interferometer gyroscope,
\textit{Phys. Rev. Lett.} \textbf{78}, 2046 (1997).

\bibitem{LenefPRL1997}
A. Lenef, T.~D. Hammond, E.~T. Smith, M.~S. Chapman, R.~A. Rubenstein, and D.~E. Pritchard,
Rotation sensing with an atom interferometer,
\textit{Phys. Rev. Lett.} \textbf{78}, 760 (1997).

\bibitem{BarrettCRP2014}
B. Barrett, R. Geiger, I. Dutta, M. Meunier, B. Canuel, A. Gauguet, P. Bouyer, and A. Landragin,
The Sagnac effect: 20 years of development in matter-wave interferometry,
\textit{C. R. Phys.} \textbf{15}, 875--883 (2014).

\bibitem{DuttaPRL2016}
I. Dutta, D. Savoie, B. Fang, B. Venon, C.~L. Garrido Alzar, R. Geiger, and A. Landragin,
Continuous cold-atom inertial sensor with 1 nrad/sec rotation stability,
\textit{Phys. Rev. Lett.} \textbf{116}, 183003 (2016).

\bibitem{GautierSCIADV2024}
R. Gautier, M. Guessoum, L.~A. Sidorenkov, Q. Bouton, A. Landragin, and R. Geiger,
Accurate measurement of the Sagnac effect for matter waves,
\textit{Sci. Adv.} \textbf{8}, eabn8009 (2022).

\bibitem{CarlosAVS2019}
C.~L. Garrido Alzar,
Compact chip-scale guided cold atom gyrometers for inertial navigation: Enabling technologies and design study,
\textit{AVS Quantum Sci.} \textbf{1}, 014702 (2019).

\bibitem{FernándezARXIV2025}
M. Frómeta Fernández, D. Hernández-Rajkov, G. Del Pace, N. Grani, M. Inguscio, F. Scazza, S. Stringari, and G. Roati,
Angular momentum of rotating fermionic superfluids by Sagnac phonon interferometry,
\textit{arXiv}:2511.02664 (2025).

\bibitem{ZhuPRA2015}
S. Zhu and B. Wu,
Interaction effects on Wannier functions for bosons in an optical lattice,
\textit{Phys. Rev. A} \textbf{92}, 063637 (2015).

\bibitem{MarzariPRB1997}
N. Marzari and D. Vanderbilt,
Maximally localized generalized Wannier functions for composite energy bands,
\textit{Phys. Rev. B} \textbf{56}, 12847 (1997).

\bibitem{footnoteSmerzi}
In the case $n=2$, the toroidal system reduces to an effective double-well trap with two tunneling barriers, where the only allowed oscillatory mode is the $k=\pi$ mode.
%
In this case, Eq.~(\ref{eq:eigenfreq_rotating}) recovers the plasma frequency of Josephson oscillations in a double-well trap~\cite{SmerziPRL1997,RaghavanPRA1999}, with the replacement $J\to 2J$ accounting for the two tunneling paths imposed by periodic boundary conditions.

\bibitem{SmerziPRL1997}
A. Smerzi, S. Fantoni, S. Giovanazzi, and S.~R. Shenoy,
Quantum coherent atomic tunneling between two trapped Bose--Einstein condensates,
\textit{Phys. Rev. Lett.} \textbf{79}, 4950 (1997).

\bibitem{RaghavanPRA1999}
S. Raghavan, A. Smerzi, S. Fantoni, and S.~R. Shenoy,
Coherent oscillations between two weakly coupled Bose--Einstein condensates: Josephson effects, $\pi$ oscillations, and macroscopic quantum self-trapping,
\textit{Phys. Rev. A} \textbf{59}, 620 (1999).

\bibitem{CataliottiSCIENCE2001}
F. S. Cataliotti, S. Burger, C. Fort, P. Maddaloni, F. Minardi, A. Trombettoni, A. Smerzi, and M. Inguscio,
Josephson junction arrays with Bose-Einstein condensates,
\textit{Science} \textbf{293}, 843 (2001).

\bibitem{AlbiezPRL2005}
M. Albiez, R. Gati, J. Fölling, S. Hunsmann, M. Cristiani, and M.~K. Oberthaler,
Direct observation of tunneling and nonlinear self-trapping in a single bosonic Josephson junction,
\textit{Phys. Rev. Lett.} \textbf{95}, 010402 (2005).

\bibitem{ValtolinaSCIENCE2015}
G. Valtolina, A. Burchianti, A. Amico, E. Neri, K. Xhani, J.~A. Seman, A. Trombettoni, A. Smerzi, M. Zaccanti, M. Inguscio, and G. Roati,
Josephson effect in fermionic superfluids across the BEC--BCS crossover,
\textit{Science} \textbf{350}, 1505 (2015).

\bibitem{SpagnolliPRL2017}
G. Spagnolli, G. Semeghini, L. Masi, G. Ferioli, A. Trenkwalder, S. Coop, M. Landini, L. Pezz\`e, G. Modugno, M. Inguscio, A. Smerzi, and M. Fattori,
Crossing over from attractive to repulsive interactions in a tunneling bosonic Josephson junction,
\textit{Phys. Rev. Lett.} \textbf{118}, 230403 (2017).

\bibitem{DelPaceSCIENCE2025}
G. Del Pace, D. Hernández-Rajkov, V.~P. Singh, N. Grani, M. Frómeta Fernández, G. Nesti, J.~A. Seman, M. Inguscio, L. Amico, and G. Roati,
Shapiro steps in strongly interacting Fermi gases,
\textit{Science} \textbf{390}, 1125 (2025).

\bibitem{BiagioniNATURE2024}
G. Biagioni, N. Antolini, B. Donelli, L. Pezz\`e, A. Smerzi, M. Fattori, A. Fioretti, C. Gabbanini, M. Inguscio, L. Tanzi, and G. Modugno,
Measurement of the superfluid fraction of a supersolid by Josephson effect,
\textit{Nature} \textbf{629}, 773 (2024).

\bibitem{DonelliPRA2025}
B. Donelli, N. Antolini, G. Biagioni, M. Fattori, A. Fioretti, C. Gabbanini, M. Inguscio, L. Tanzi, G. Modugno, A. Smerzi, and L. Pezz\`e,
Self-induced Josephson oscillations and self-trapping in a supersolid dipolar quantum gas,
\textit{Phys. Rev. A} \textbf{112}, L051302 (2025).

\bibitem{TengstrandPRA2021}
M.~N. Tengstrand, D. Boholm, R. Sachdeva, J. Bengtsson, and S.~M. Reimann,
Persistent currents in toroidal dipolar supersolids,
\textit{Phys. Rev. A} \textbf{103}, 013313 (2021).

\bibitem{TengstrandPRA2023}
M. N. Tengstrand, P. Stürmer, J. Ribbing, and S. M. Reimann.
Toroidal dipolar supersolid with a rotating weak link
\textit{Phys. Rev. A} \textbf{107}, 063316 (2023).

\bibitem{ŠindikPRL2024}
M. Šindik, T. Zawiślak, A. Recati, and S. Stringari,
Sound, Superfluidity, and Layer Compressibility in a Ring Dipolar Supersolid
\textit{Phys. Rev. Lett.} \textbf{132}, 146001 (2024).

\bibitem{PretiPRL2026}
N. Preti, N. Antolini, C. Drevon, P. Lombardi, A. Fioretti, C. Gabbanini, G. Ferioli, G. Modugno, and G. Biagioni,
Single-fluid model for rotating annular supersolids and its experimental implications,
\textit{Phys. Rev. Lett.} \textbf{136}, 036001 (2026).

\bibitem{VoronovaNATCOMM2025}
N. Voronova, A. Grudinina, R. Panico, D. Trypogeorgos, M. De Giorgi, K. Baldwin, L. Pfeiffer, D. Sanvitto, and D. Ballarini,
Exciton-polariton ring Josephson junction,
\textit{Nat. Commun.} \textbf{16}, 466 (2025).

\bibitem{GatiJPB2007}
R. Gati and M.~K. Oberthaler,
A bosonic Josephson junction,
\textit{J. Phys. B: At. Mol. Opt. Phys.} \textbf{40}, R61--R89 (2007).

\bibitem{EstèveNATURE2008}
J. Estève, C. Gross, A. Weller, S. Giovanazzi, and M.~K. Oberthaler,
Squeezing and entanglement in a Bose--Einstein condensate,
\textit{Nature} \textbf{455}, 1216--1219 (2008).

\bibitem{BerradaNATCOMM2013}
T. Berrada, S. van Frank, R. Bücker, T. Schumm, J.-F. Schaff, and J. Schmiedmayer,
Integrated Mach--Zehnder interferometer for Bose--Einstein condensates,
\textit{Nat. Commun.} \textbf{4}, 2077 (2013).

\bibitem{PezzeRMP2018}
L. Pezz\`e, A. Smerzi, M.~K. Oberthaler, R. Schmied, and P. Treutlein,
Quantum metrology with nonclassical states of atomic ensembles,
\textit{Rev. Mod. Phys.} \textbf{90}, 035005 (2018).


\end{thebibliography}
\end{document}